\documentclass[11pt]{article}
\pdfoutput=1

\usepackage[]{Packages}
\bibliography{T-duality}

\usepackage{Definitions}

\newcommand*{\lattice}{
    \coordinate (Origin)   at (0,0);
    \coordinate (XAxisMin) at (-1,0);
    \coordinate (XAxisMax) at (6,0);
    \coordinate (YAxisMin) at (0,-1);
    \coordinate (YAxisMax) at (0,6);
    \draw [thin, gray,-latex] (XAxisMin) -- (XAxisMax);%
    \draw [thin, gray,-latex] (YAxisMin) -- (YAxisMax);%

    \clip (-1,-1) rectangle (6,6); %

    \draw[style=help lines,dashed] (-2,-2) grid[step=1] (7,7);

    \pgftransformcm{\coeffk}{0}{\coeffi}{\coeffkp}{\pgfpoint{0}{0}}

    \coordinate (Xone) at (1,0);
    \coordinate (Xtwo) at (0,1);

    \foreach \x in {-7,-6,...,7}{%
      \foreach \y in {-7,-6,...,7}{%
        \node[draw,circle,inner sep=2pt,fill] at (1*\x,1*\y) {};
      }
    }
    \filldraw[fill=gray, fill opacity=0.3, draw=black] (Origin)
        rectangle ($(Xone)+(Xtwo)$);

    \draw[style=help lines,dashed, red] (-2,-2) grid[step=1] (7,7);

    \draw [thick,-latex,red] (Origin) -- (Xone) node [below] {$x_1$};
    \draw [thick,-latex,red] (Origin) -- (Xtwo) node [above] {$x_2$};
}

\preprint{\texttt{}}

\newcommand{\OfficialTitle}{\huge{The M--theory origin of global properties of gauge theories}}

\title{\vspace{2cm}
  {\Huge\textbf{\dosserif\OfficialTitle}}
}

\hypersetup{pdfauthor={Antonio Amariti and Claudius Klare and Domenico Orlando and Diego Redigolo and Susanne Reffert},pdftitle={\OfficialTitle}}

\author{%
  \begin{minipage}{.8\linewidth}
    \vspace{1cm}
    \begin{center} \dosserif
      {\small 
        \textbf{Antonio~Amariti}\textsuperscript{$\spadesuit$},
        \textbf{Claudius~Klare}\textsuperscript{$\heartsuit$}, 
        \textbf{Domenico~Orlando}\textsuperscript{$\spadesuit,\clubsuit$}, 
         and 
        \textbf{Susanne~Reffert}\textsuperscript{$\diamondsuit$}}
    \end{center}
    \vspace{1cm}
       \authorBlock{$\spadesuit$}{\textsc{lptens} -- \textsc{umr cnrs} 8549, 24, rue Lhomond, F-75231 Paris, France} 
    \authorBlock{$\heartsuit$}{\textsc{ipht}, \textsc{cea}/Saclay, F-91191 Gif-sur-Yvette, France}
    \authorBlock{$\clubsuit$}{\textsc{ipt} Ph. Meyer, 24, rue Lhomond, F-75231 Paris, France}
    \authorBlock{$\diamondsuit$}{\textsc{itp -- aec}, University of Bern, Sidlerstrasse 5, CH-3012 Bern, Switzerland}
  \end{minipage}
}

\date{}

\begin{document}

\setstretch{1.15}

\numberwithin{equation}{section}

\begin{titlepage}

  \newgeometry{top=23.1mm,bottom=46.1mm,left=34.6mm,right=34.6mm}

  \maketitle

  \thispagestyle{empty}

  \vfill\dosserif
  \vspace{1cm}
  \abstract{\normalfont \noindent
    We show that global properties of gauge groups can be understood as geometric properties in M--theory.
    Different wrappings of a system of \(N\) \M5--branes on a torus reduce to four-dimensional theories with \(A_{N-1}\) gauge algebra and different unitary groups. The classical properties of the wrappings determine the global properties of the gauge theories without the need to impose any quantum conditions.  We count the inequivalent wrappings as they fall into orbits of the modular group of the torus, which correspond to the S-duality orbits of the gauge theories.
  }

  \vfill

\end{titlepage}

\restoregeometry

\section{Introduction}

In this work, we show how global quantum properties of gauge theories are determined purely by the classical set-up of its M--theory parent theory. We consider $N$ \M5--branes wrapped on a two-torus. We probe this geometry using \M2--branes wrapping a curve on the two-torus. The geometrical requirement that this curve is closed translates directly into the quantization condition of \acl{dsz} for dyonic lines after reduction to four-dimensional field theory, without the need of imposing any external constraint. This is yet another example of the power of the M--theory point of view which allows us to make quantum statements for gauge theories based on purely classical considerations in M--theory, see \emph{e.g.}~\cite{Witten:1997sc,Bachas:1997sc}. Our results tie in with a number of recent field theory results which study line operators in gauge theory~\cite{Kapustin:2005py,Gomis:2006im,Kapustin:2007wm,Drukker:2009tz,Gaiotto:2010be,Aharony:2013hda, Xie:2013vfa,Xie:2013lca,Bullimore:2013xsa,Gaiotto:2014kfa,Moore:2014gua,Okuda:2014fja,Coman:2015lna}.

\bigskip
In the analysis of Abelian gauge theories, there are two types of charges associated to the
gauge group, electric and magnetic. These charges are 
constrained by the Dirac quantization condition:
an electrically charged particle living in the field of a monopole
must have a quantized charge in terms of units of the Planck's constant.
A similar condition must be imposed when there are particles charged both electrically and magnetically, \emph{i.e.} dyons.
In this case the condition involves pairs of charges and is known as the \ac{dsz} condition.
One can consider also non-Abelian \ac{ym} gauge theories. In this case, 
the magnetic charges are the \ac{gno} charges.
The same quantization condition exists for this case and involves the 
charges of the fields under the center of the gauge group.
It is important to observe that the quantization condition exists only for quantum theories: the condition 
follows after imposing some extra constraint, for example 
the quantization of the angular momentum.
 
The \ac{dsz} condition has been recently used in~\cite{Aharony:2013hda} to read off the global properties of the gauge group associated to 
a gauge algebra.
When considering \ac{ym} theories one can study the spectrum of \acp{wline}
under different representations of the algebra. These lines are the electrically charged objects. 
The magnetically charged objects are the \acp{hline}.
For fixed gauge algebra, that the gauge group is specified by the spectrum of representations of the allowed \acp{wline}, \acp{hline} or \acp{hwline}.
The final result is that a group is specified by the maximal choice of mutually local lines. The requirement of 
mutual locality corresponds to the validity %
of the \ac{dsz} quantization condition between each pair of lines. 
 
An interesting result following from this analysis regards
$\mathcal{N}=4$ \ac{sym} theories. In this case, the distinct gauge groups
are naively related to each other by S--duality~\cite{Montonen:1977sn}, or more mathematically 
by the action of the modular group
$SL(2,\mathbb{Z})$. After a careful analysis of the global properties 
it has been observed that the situation is more involved.
There are indeed \emph{orbits} under the modular group
and the action of $SL(2,\mathbb{Z})$ cannot link all the theories 
with the same gauge algebra.

Motivated by these field theory results, we analyze the problem in this paper from the 
M--theory perspective. 
Our two main results are the following.
\begin{itemize}
\item
We show that the structure of $SL(2,\mathbb{Z})$ depends on the 
torus defined by wrapping $N$ \M5--branes along the two compact directions of M--theory.
There are indeed many possible tori defined by the wrapping of the \M5--branes, 
and they correspond to the choice of the gauge group.
In general the new tori are not all mapped into each other by modular transformations.    
This corresponds to the existence of orbits in the action of $SL(2,\mathbb{Z})$ in $\mathcal{N}=4$ \ac{sym}.
\item
We give a geometric derivation of the \ac{dsz} condition without any reference to the quantum properties:
a quantum condition on the field theory side becomes geometrical and classical in M--theory.
\end{itemize}
The plan of our note is as follows. We begin our analysis with the geometric description of $N$ \M5--branes wrapping the torus of
M--theory in Section~\ref{sec:Msetup}. This torus defines a  $\mathbb{Z}^2$ 
lattice. When we consider $N$ \M5--branes wrapping this torus we have
to fix a choice of $N$ cells in this lattice. This choice fixes a
sublattice of index $N$ (\emph{i.e.} a lattice in \(\setZ^2\) where the elementary cell has area \(N\)) and this sublattice
defines a new torus that we will call $T_{N;k,i}$. Note that the restriction of a sublattice \(\Lambda_{N;k,i} \subset \setZ^2\) defines a Lagrangian (maximal) sublattice in \(\setZ_N \times \setZ_N\): this condition is not imposed but is automatically satisfied by our setup for the \M5--brane embedding.
There is a geometric way to probe $T_{N;k,i}$, it consists of wrapping \M2--branes along closed curves on $T_{N;k,i}$.
The intersection number of pairs of \M2--branes corresponds to the \ac{dsz} condition.

In order to associate this geometric construction to the global properties of gauge theories 
we first reduce the M--theory to \tIIA and then T--dualize along the remaining compact direction, see Section~\ref{sec:MtoString}. In this way 
we obtain a \tIIB description with a stack of $N$ parallel \D3--branes representing 4d $\mathcal{N}=4$ \ac{sym} with gauge algebra $su(N)$. 
The \M2--lines become bound states of \D1/\F1 strings. They correspond to \acp{hwline}.
The set of allowed bound states depends on the geometric construction in M--theory.
We can associate a set of bound states to each $T_{N;k,i}$, 
\emph{i.e.} to a lattice. Eventually $T_{N;k,i}$ lattices correspond to the 
lattices of allowed charges of the \D1/\F1   bound states.
We find in Section~\ref{sec:field} that these correspond to the lattices discussed in~\cite{Aharony:2013hda} and fix the global properties as
anticipated. For clarity, the explicit example of the case of $SU(4)$ is discussed in Section~\ref{sec:example}.

In Section~\ref{sec:Open}, we conclude with a comparison to field theory results and open questions.
In Appendix~\ref{app:Counting}, we derive the number of orbits of \(SL(2, \setZ)\) in the set \(\Gamma(N)\) of sublattices of index \(N\).
In Appendix~\ref{app:Lie}, we discuss the representations of the $A_{N-1}$ algebra in terms of \D3/\D1 systems.

\section{M--theory setup}\label{sec:Msetup}

\subsection{The torus $T_{N;k,i}$}

\paragraph{M5--branes and sublattices.}

We consider $N$ \M5--branes extended in the directions $\set{x^0,x^1,x^2,x^3}$ and wrapped on a torus $(x^9,x^{10})$. The torus is identified by a fundamental domain with sides \(y_1, y_2 \in \setC\). Equivalently, this defines the $\mathbb{Z}^2$--lattice generated by $y_1$ and $y_2$. Wrapping the \M5--branes corresponds to choose $N$ cells in this lattice, \emph{i.e.} defining a sublattice of index $N$. The standard way of defining a sublattice is to work in the Hermite normal form. Let $\langle y_1,\dots y_d\rangle$ be the generators of a lattice. Any sublattice is generated by $\langle x_1,\dots x_d\rangle$, which can always be written in the form
\begin{equation}
  \begin{cases}
    x_1 = a_{11}y_1,\\
    x_2 = a_{21}y_1+a_{22}y_2,\\
    \vdots \\
    x_d = a_{d1}y_1+\dots+a_{dd}y_d,
  \end{cases}
\end{equation}
where the positive integers $a_{ij}$ satisfy $0\leq a_{ij}<a_{jj}$ for all $i > j$ and the index is $N=\prod_i a_{ii}$.
In our case, $d=2$ and we set $a_{11}=k,\, a_{21}=i,\, a_{22}=k'$ such that
\begin{equation}
  \begin{cases}
    x_1 = k y_1,\\
    x_2 = i y_1 + k'y_2,
  \end{cases}
\end{equation}
where $kk'=N$ and $0\leq i <k$. We denote the sublattice generated by
\(\langle x_1, x_2 \rangle\) as \( \Gamma_{N;k,i} \subset \Gamma_{1;1,0} = \setZ^2\). Equivalently, \(\Gamma_{N;k,i}\) defines a new torus \(T_{N;k,i}\) which contains all the information about the wrapping of the branes.

\paragraph{Counting and generating functions.}

It is immediate to see that for fixed $N$, there are $f(N)=\sum_{k|N} k$ sublattices (the sum runs over all the divisors of \(N\)). This information can be encoded in a generating function using the fact that $f$ is multiplicative and decomposable as a convolution (see Appendix~\ref{app:Counting} for details):
\begin{equation}
  f(N) = \sum_{m|N} u(m) N(N/m) = (u\star N)(N),
\end{equation}
where $u(N)=1$ for all $N$ and $N(N)=N$.
It follows that
\begin{align}
  &\begin{aligned}
    \mathcal{F}(t) &\coloneqq \sum_{N=1}^\infty f(N)t^N = \sum_{j=1}^\infty \sum_{k=1}^\infty k\, t^{jk}\\
    &= \sum_{j=1}^\infty \frac{1+t^{3j}}{(1-t^j)(1- t^{2j})} - 1 =
    \sum_{j=1}^\infty \sum_{n_1, n_2=0}^\infty (1+t^{3j}) t^{j(n_1+2n_2)} - 1,
  \end{aligned} \\
  &\mathscr{F}(s) \coloneqq \sum_{N=1}^\infty \frac{f(N)}{N^s} = \zeta(s) \zeta(s-1) = \prod_p \frac{1}{1- p^{-s} - p^{-s + 1} + p^{-2s + 1}},
\end{align}
where $\zeta(s)$ is the Riemann $\zeta$--function.
From the form of the Dirichlet series we derive the average number of sublattices:
\begin{equation}
  \frac{1}{\bar N} \sum_{N = 1}^{\bar N} f(N) \xrightarrow[\bar N \to \infty]{} \frac{\pi^2}{12} \bar N .
\end{equation}

\bigskip

\begin{figure}
  \centering
  \begin{tikzpicture}[x=4cm, y=4cm]

    \foreach \name / \Xcoord / \Ycoord / \coeffk / \coeffkp / \coeffi in {10/0/2/1/4/0, 40/0/1/4/1/0, 41/-1/0/4/1/1, 43/1/0/4/1/3, 42/0/-1/4/1/2, 21/0/-2/2/2/1, 20/1.5/1.5/2/2/0} {
      \node (\name) at (\Xcoord, \Ycoord) {
        \begin{tikzpicture}[x=.4cm, y=.4cm]
          \lattice
        \end{tikzpicture}
      };
      \draw (\name) ++(.5,.3) node {\(\Gamma_{4;\coeffk,\coeffi}\)};
    }

    \draw [latex-latex] (10) -- (40) node [midway,right] {\(S\)};
    \draw [latex-latex] (41) -- (43) node [midway,above] {\(S\)};
    \draw [latex-latex] (42) -- (21) node [midway,right] {\(S\)};

    \draw [-latex] (40) -- (41) node [midway,above] {\(T\)};
    \draw [-latex] (41) -- (42) node [midway,above] {\(T\)};
    \draw [-latex] (42) -- (43) node [midway,above] {\(T\)};
    \draw [-latex] (43) -- (40) node [midway,above] {\(T\)};

    \draw [-latex] (10) ++(-.4,0) arc[x radius =.1, y radius =.1, start angle=30, end angle=330];
    \node at (-.5,2.1){\(T\)};
    \draw [-latex] (21) ++(.4,0) arc[x radius =.1, y radius =.1, start angle=150, end angle=-150];
    \node at (.5,-1.9){\(T\)};

    \draw [-latex] (20) ++(.4,0) arc[x radius =.1, y radius =.1, start angle=150, end angle=-150];
    \node at (2,1.6){\(S,T\)};
  \end{tikzpicture}
  \caption{Index \(4\) sublattices of \(\setZ^2\) (the set \(\Gamma(4)\)) represented as the \(2\) orbits under modular transformations.}
  \label{fig:4-M5-lattices}
\end{figure}
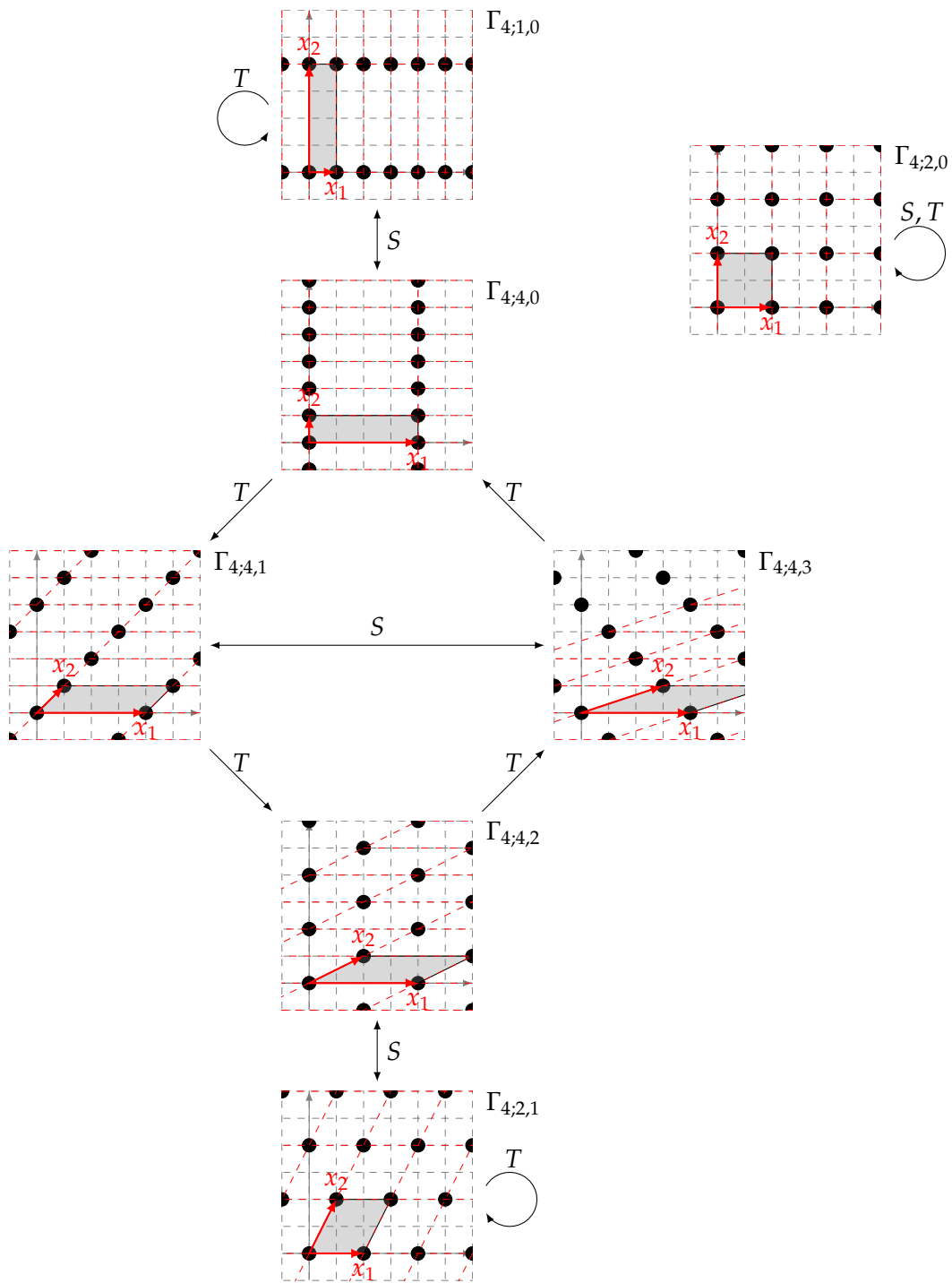

From a purely geometrical point of view, the sublattices
\(\Gamma_{N;k,i}\) are not inequivalent: some of them can be related
via an $SL(2,\mathbb{Z})$ transformation which will in general change
\(k\) and \(i\) and leave \(N\) invariant
(Figure~\ref{fig:4-M5-lattices}). We can define a new sequence $\hat
f(N)$ counting the number of sublattices up to the $SL(2,\mathbb{Z})$
action. This sequence is again multiplicative and, as it is usually convenient in counting sublattices~\cite{Hanany:2010cx}, can be decomposed in terms of a convolution of \(u\):
\begin{equation}
\hat f(N)=(u\star sq)(N),
\end{equation}
where $sq(N)$ is the characteristic function of squares given by
\begin{align}
sq(N) = 
  \begin{cases}
    1 & \text{if \(\exists  b \in \mathbb{N}:\ N=b^2\),}\\
    0 & \text{otherwise.}
  \end{cases}
\end{align}
(See Appendix~\ref{app:Counting} for a proof).
It follows immediately that the generating functions are
\begin{align}
  \hat {\mathcal{F}}(t) &\coloneqq \sum_{N=1}^\infty \hat f(N) t^N = \sum_{a=1}^\infty \sum_{b=1}^\infty t^{ab^2} = \sum_{k=1}^\infty \frac{1}{1-t^{k^2}} - 1, \\
  \hat {\mathscr{F}}(s) &\coloneqq \sum_{N=1}^\infty \frac{\hat f(N)}{N^s} = \zeta(s) \zeta(2s) = \prod_p \frac{1}{1 - p^{-s} - p^{-2s} + p^{-3s}}.
\end{align}
In this case the average number of sublattices is asymptotically constant:
\begin{equation}
  \frac{1}{\bar N} \sum_{N = 1}^{\bar N} \hat f(N) \xrightarrow[\bar N \to \infty]{} \frac{\pi^2}{6} .
\end{equation}

From the expression of \(\hat {\mathcal{F}} (t)\) we see that $\hat f(N)$ counts the number of times
in which \(N\) is divisible by a perfect square. Following the construction in Appendix~\ref{app:Counting}, each orbit has a representative of the form
\begin{equation}
  O_{a,b} =
  \begin{pmatrix}
    a b& 0 \\ 0 & b
  \end{pmatrix}, \quad ab^2 = N .
\end{equation}

\subsection{M2s wrapping $T_{N;k,i}$: the DSZ condition}

Consider now an \M2--brane extended in \(x^0, x^4\) and wrapping a geodesic \(\mathcal{C}\) on the torus \(T_{N;k,i}\).\footnote{The \M5/\M2 system preserves a total of eight supercharges.}
Since there are no other objects in the theory where the curve could end, \(\mathcal{C}\) must be closed. The only constraint that we impose is that the curve does not cover the torus. %
In other words, we demand the curve to have homology \((p,q)_{N;k,i}\) with respect to the cycles of the torus \(T_{N;k,i}\). 

\begin{figure}
  \centering
  \begin{tikzpicture}
    \newcommand{\coeffk}{2}    
    \newcommand{\coeffi}{1}
    \newcommand{\coeffkp}{2}
    
    \begin{scope}[x=.7cm, y=.7cm]
      \lattice
      \draw [thick,-latex,blue] (Origin) -- (2,1) node [below] {$(5,2)$};
    \end{scope}

    \begin{scope}[x=1cm, y=1cm, xshift=9cm, yshift=2cm]
      \node at (0,0) {\includegraphics[width=.45\textwidth]{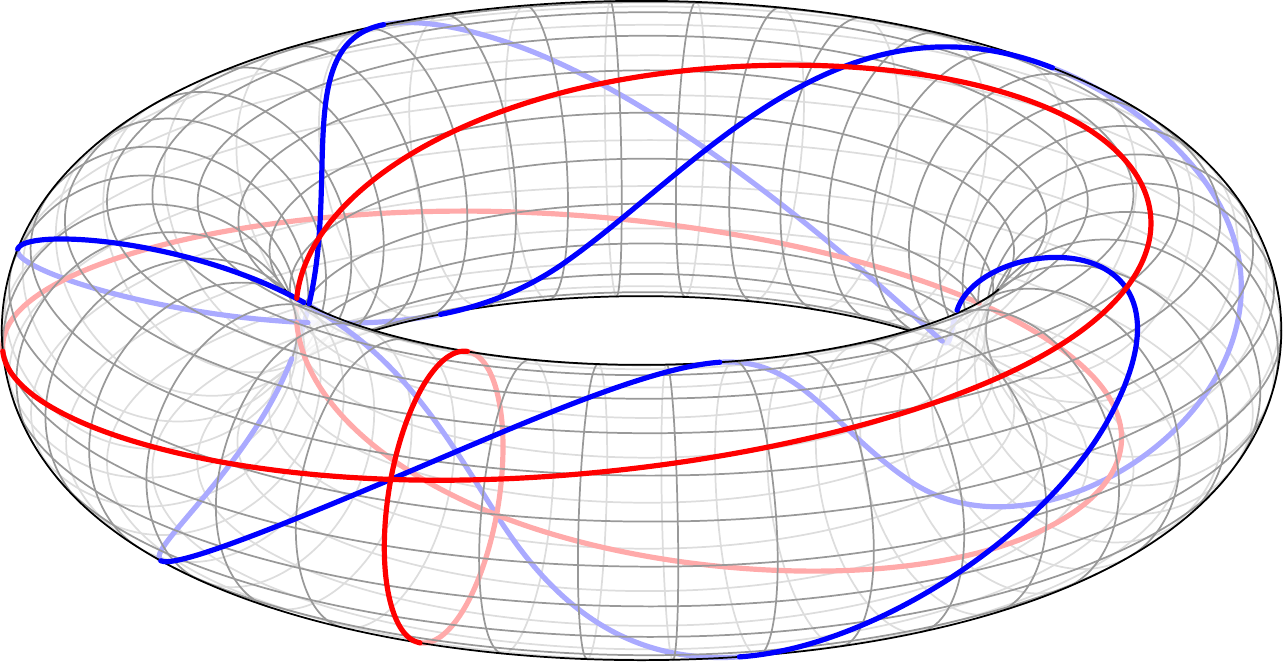}};
    \end{scope}

  \end{tikzpicture}
  \caption{In blue the \((5,2)\) curve as a vector in the lattice
    \(\Gamma_{4;2,1}\) (left) and as a line on the torus \(T\) (right). The line has homology \((e=5,m=2) = (p=2,q=1)_{4;2,1}\). In red the generators \(x_1 = (2,0)\) and \(x_2 = (1,2)\).
    After the reduction to \tIIB, this curve will correspond to a dyon of charge \((5,2)\) in \((SU(4)/\setZ_2)_1\) (Section~\ref{sec:field}).}
  \label{fig:5-2-dyon}
\end{figure}

In turn, since \(x_1 = k y_1\) and \(x_2 = i y_1 + k' y_2\), the curve \(\mathcal{C}\) has homology \((e,m) = (p k + q i, q k')\) with respect to the torus \(T\) (remember that \(T_{N;k,i}\) is an \(N\)--cover of \(T\)). The curve can be represented by a vector in the plane joining the origin to the point \((pk + qi, q k')\) (Figure~\ref{fig:5-2-dyon}).
Using the homology class, we can evaluate the intersection number of two curves \(\mathcal{C}\) and \(\mathcal{C}'\) in terms of the symplectic structures on \(T\) (or equivalently \(T_{N;k,i}\)):
\begin{equation}
\label{eq:intersection}
  \begin{aligned}
    \# (\mathcal{C}, \mathcal{C}') &=
    \begin{pmatrix}
      p k + q i & q k'
    \end{pmatrix}
    \begin{pmatrix}
      0 & 1 \\ -1 & 0
    \end{pmatrix}
    \begin{pmatrix}
      p' k + q' i \\ q' k'
    \end{pmatrix} \\
    &=
    \begin{pmatrix}
      p & q
    \end{pmatrix}
    \begin{pmatrix}
      0 & N \\ -N & 0
    \end{pmatrix}
    \begin{pmatrix}
      p' \\ q'
    \end{pmatrix}
    = \left( p q' - q p' \right) N,
  \end{aligned}
\end{equation}
where we used the fact that \(k k' = N\).

It is worth to stopping a moment to stress this result. By construction, the \M2--branes correspond to curves that have to be closed and we only impose a geometrical condition (\emph{i.e.} that they do not fill the torus). We find that such curves intersect in a number of points that satisfies the condition \(\#(\mathcal{C}, \mathcal{C}') = 0 \mod N\). In the following we will see how this classical condition in M--theory translates into the quantum \ac{dsz} condition in gauge theory.

\section{From M--theory to string theory}\label{sec:MtoString}

In this section we discuss the reduction of the M--theory setup to \tIIB.
This reduction is necessary to link the lattices obtained in M--theory 
to the global properties of the gauge group in field theory. We show that the
\M2--branes become bound states of  \D1/\F1 branes in \tIIB. These bound states 
play the role of the \acp{wline} and  \acp{hline} in field theory. The M--theory lattices 
become lattices of charges of these lines and the global properties can be read from these lattices.

We reduce the M--theory configuration to \tIIB by reducing on \(x^{10} = 2 \pi R_{10} y_1\) and T--dualizing on 
\(x^9 = 2 \pi R_9 y_2\). We consider the decompactification limit, such that the T--dual radius  $\widetilde R_9$ becomes non-compact.
The \M5--branes turn into \D3--branes and an \M2--brane wrapped on a curve \(\mathcal{C}\) of homology \((p,q)_{N;k,i}\) turns into a bound state of \(e = (pk+q i)\) \F1 and \(m = qk'\) \D1--branes. 
Equation \eqref{eq:intersection} reduces to
\begin{equation}\label{eq:dzs}
  e m'-m e'=0 \mod N    
\end{equation}
for a pair of bound states $(e,m)$ and $(e',m')$.
The resulting configuration is summarized in Table~\ref{tab:IIB_branes}.
\begin{table}
  \centering
  \begin{tabular}{lcccccccccc}
    \toprule
        & 0        & 1        & 2        & 3        & 4        & 5 & 6 & 7 & 8 & 9 \\ \midrule
    \D3 & $\times$ & $\times$ & $\times$ & $\times$ &          &   &   &   &   &   \\
    \D1 & $\times$ &          &          &          & $\times$ &   &   &   &   &   \\
    \F1 & $\times$ &          &          &          & $\times$ &   &   &   &   &   \\ \bottomrule
  \end{tabular}
  \caption{Brane setup after reduction from M--theory to \tIIB}
  \label{tab:IIB_branes}
\end{table}

Each choice of the embedding of the \M5--branes corresponds to a sublattice \(\Gamma_{N;k,i}\) which here turns into a configuration of \D3--branes which we probe with the \D1/\F1 bound states. Two sublattices \(\Gamma_{N; k,i}\) and \(\Gamma_{N; k', i'}\), related by an \(SL(2, \setZ)\) transformation, correspond at this level to two \tIIB configurations related by S--duality.

The way in which the \M5--branes wrap the torus fixes the theory completely. At the string theory level we lose the geometric picture since we have reduced on the torus and it becomes convenient to introduce the probe \F1/\D1 branes to distinguish among parallel \D3--brane configurations which source different combinations of NS and RR fluxes.
\F1--branes are associated to \acp{wline} while \D1--branes are associated to \acp{hline}.

When considering the \tIIB setup, there are two kinds of \F1 and \D1--branes. There are zero-length strings connecting two \D3s in the stack, and there are
semi-infinite \F1 and \D1--branes extended in the direction $x^4$, with one endpoint ending at infinity and the other on the stack. These are respectively regular and singular monopoles.
The latter configurations have been studied in~\cite{Moore:2014jfa,Moore:2014gua} (see also~\cite{Gomis:2006im}).
By allowing the presence of semi-infinite branes ending on the stack of \D3--branes, we can systematically study a generic representation of the algebra in terms of the associated \acp{wline} and  \acp{hline}.
In Appendix~\ref{app:Lie}, we discuss the representation theory of these systems.

It turns out that we can associate the lattices obtained in M--theory to the spectrum of \D1/\F1 bound states.
The connection works as follows.
First we consider a bound state with $m$ \D1--strings (the same discussion holds in the case of $e$ \F1--branes). 
We can associate to this configuration
the $m$-index symmetric representation for the \ac{hline}. Other representations can be constructed acting with the roots which is equivalent to adding zero-length \D1--strings connecting pairs of \D3--branes in the stack (see Appendix~\ref{app:Lie}).

In general this procedure 
gives a descendant in the weight decomposition of the 
$m$-index symmetric representation.
Observe that some of the descendants coincide with the highest weight of
lower dimensional representations
carrying the same number of semi-infinite branes.
In this way we construct a sublattice of the weight lattice, such that each point 
is associated to one or more representations, identified by the same 
number of semi-infinite \D1--branes (containing the same number of boxes in the Young tableau). 
We project the weight lattice to this one-dimensional lattice and choose a representative of each point. Here we choose 
the $m$-index symmetric representation.

By repeating the discussion for the \F1--branes we obtain a two-dimensional lattice. This lattice coincides with
the lattice \(\Lambda_{N;k,i}\) obtained from the M--theory construction. It is in general defined by two points $(k,0)$
and $(i,k')$. Indeed the integers $k$, $k'$ and $i$ coincide with the number of semi-infinite \F1 and \D1--branes.

We can extract the general rule to construct the lattices in the \tIIB description by choosing the two generating points and associating them to two brane configurations. The point $(k,0)$  corresponds to the $k$-index symmetric representation, \emph{i.e.} a stack of $k$ semi--infinite \F1--strings ending on the \(N\)-th \D3--brane on the stack. The point $(i,k')$ is a stack of  $i$ \F1--strings and $k'$ \D1--branes ending on the \(N\)-th \D3--brane.
These two configurations generate all the possible representations: by acting with the roots (adding finite branes in the stack of \D3s) we obtain the lower-dimensional representations with the same number of boxes. Other representations associated to the lattice are obtained by linearly combining the lines $(k,0)$ and $(i,k')$.

\section{Field theory interpretation}\label{sec:field}

We can now interpret the lattices associated to the
bound states of \F1/\D1 strings in terms of field theory.
As already observed in Section~\ref{sec:MtoString}, these lattices coincide with the charges of the \acp{wline} and \acp{hline}.
The set of allowed charges of these line operators specifies the 
global properties of the gauge group~\cite{Aharony:2013dha}.
At the level of the Lagrangian, the gauge group is fixed by specifying the $\theta$-angle, its periodicity and some possible discrete shift. These properties do not affect the correlators but they can be probed by putting the theory on a spin manifold~\cite{Vafa:1994tf}\footnote{This idea has been used in~\cite{Razamat:2013opa} for studying 
the global properties on the superconformal index on a Lens space.
} or, alternatively, as discussed here, by probing the theory with \textsc{hw} operators.

In this section we first review the argument of~\cite{Aharony:2013dha} and show that
 our M--theory construction coincides with the classification of the different global 
groups associated to the $su(N)$ algebra.
In other words, we show that the global properties of the gauge theory are specified by the choice of the torus \(T_{N;k,i}\) that encodes the wrapping of the \M5--branes.

\subsection{Gauge groups from W and H lines}

Consider a connected gauge group $G$ 
associated to a Lie algebra $g$.
Fixing $G$ from $g$ requires some extra information because  
there are different possible gauge groups corresponding to the same algebra. 
The gauge group is fixed as follows:  consider the universal covering group $\widetilde G$ and mod it out 
 by its center $\mathbf{C}$ or a subgroup $\mathbf{H} \subset \mathbf{C}$. 
The gauge group is $G=\widetilde G/\mathbf{H}$.
This group is commonly referred to as the electric gauge group. 
Fixing completely the global properties of the gauge group requires a similar discussion on the magnetic side.
The magnetic (\ac{gno}) dual group $G^*$ is not completely fixed by the electric choice $G=\widetilde G/\mathbf{H}$.
The dual group is fixed by a choice of the charges in the \ac{gno}--dual $g^*$ algebra.

A possible way to fix the gauge group consists of choosing a (maximal) set of allowed line operators.
These line operators are \acp{wline} and \acp{hline}. The \acp{wline} fix the the global properties of the electric group and the \acp{hline} are used to fix the properties of the magnetic gauge group.

In 4d $\mathcal{N}=4$ $su(N)$  \ac{sym} we consider \acp{wline} preserving half of the supersymmetry. They are  
called $1/2$ \ac{bps} \acp{wline} and are defined as
\begin{equation}
  W_R = \Tr_R P \exp \left[ \int (i A_0 + \phi) \di t \right],
\end{equation}
where $R$ corresponds to an irreducible representation of $su(N)$.
The scalar $\phi$ corresponds to one of the six scalars in the  $\mathcal{N}=4$  vector multiplet.
These operators are the electrically charged operators.

The magnetic lines correspond to $1/2$ \ac{bps} \textsc{h}--operators. They are the S--dual
of the \acp{wline}.
An \ac{hline} corresponds to the insertion of a Dirac monopole of charge $m$ at the origin~\cite{'tHooft:1981ht}. %
The boundary behavior of $F$ and $\phi$ in spherical coordinates is
\begin{align}
F &\simeq \frac{m}{2} \sin \theta d \theta \wedge d\varphi, &
\phi &\simeq - \frac{m}{2 r}.
\end{align}

In general, both \acp{wline} and \acp{hline} preserve half of the original supersymmetry.
One can also consider more general 1/2 \ac{bps} \acp{hwline}, preserving a different half
of the original supersymmetry.

After we fix the group  $G=\widetilde G/\mathbf{H}$,
the line operators have to be invariant under $\mathbf{H}$.
Here \acp{wline} are invariant under $\mathbf{H}$, \emph{i.e.} they are labeled by representations
of $G$. Analogously, the \acp{hline} are labeled by representations of $G^*$.
The spectrum of the allowed operators is in one-to-one correspondence with the 
gauge group.
This problem can be further simplified: it is not necessary to specify all the possible representations
but one can restrict to a subset of them, identified by their charge under the center $\mathbf{C}$.

Let us first discuss the case of the electric lines, the  \acp{wline}.
If the gauge group is the universal cover $\widetilde G$ there is no restriction on the allowed representations.
We can associate the representations to a lattice, the weight lattice $\Lambda_w$ of $g$, modulo the Weyl group $W$.
If we mod $\widetilde G$ by $\mathbf{C}$ only few representations survive. These are the representations invariant under $\mathbf{C}$.
In terms of lattices, the allowed representations are obtained by acting with the roots of $g$
on a sublattice specified by the adjoint representation. The weights of the adjoint are the roots of the algebra $g$ and this sublattice is denoted as the root lattice $\Lambda_r$.
One can choose also a subgroup $\mathbf{H} \subset \mathbf{C}$. In this case, the representation live in a sublattice called the co-character lattice $\Gamma_G^*$.
The three lattices are related as $\Lambda_c \subset \Gamma_G^* \subset \Lambda_w$.
The center $\mathbf{H}$ and the fundamental group $\pi_1(G)$ are
\begin{eqnarray}
\mathbf{H} = \Gamma_G^*/\Lambda_r,
\quad
\pi_1(G) = \Lambda_{w}/\Gamma_G^*.
\end{eqnarray}

A similar discussion can be applied to the \ac{gno}--dual algebra $g^*$. The center is $\mathbf{C^*} = \mathbf{C}$
and the Weyl group is $W^*=W$.
The magnetic weight lattice $\Lambda_{mw}$ corresponds to the dual of the root lattice of $g$.
 The magnetic root lattice is called $\Lambda_{cr}$, and it is usually called the co-root lattice.
 The dual of the co-character lattice is called the  character lattice $\Gamma_{G^*}$. The inclusion here is
 $\Lambda_{cr} \subset \Gamma_{G^*} \subset \Lambda_{mw}$.
The center $\mathbf{H}$ and the fundamental group $\pi_1(G^*)$ are
\begin{eqnarray}
\mathbf{H^*} =  \Lambda_{mw}/\Gamma_{G^*},
\quad
\pi_1(G^*) = \Gamma_{G^*}/ \Lambda_{cr}.
\end{eqnarray}
In the magnetic case, the set of lines corresponds to the 't Hooft lines.
The gauge group is fixed once their allowed representations are specified.

Observe that there are not only purely electric \acp{wline} or purely magnetic \acp{hline}
but we can allow also dyonic \acp{hwline}.
It turns out that specifying the set of allowed \acp{wline} and \acp{hline} corresponds to specify 
a sublattice  of $ \Lambda_w \times \Lambda_{mw}$ modulo the action of the Weyl group.

Consider a general dyonic line $(\lambda_e,\lambda_m) \in  \Lambda_w \times \Lambda_{mw}$.
It is identified with the line $(w \lambda_e,w \lambda_m)$, where $w$ is an element of $W$.
If the line  $(\lambda_e,\lambda_m)$ is allowed, also the line  $(-\lambda_e,-\lambda_m)$ is allowed.
Moreover if two lines $(\lambda_e,\lambda_m)$  and $(\lambda'_e,\lambda'_m)$ are allowed, 
also the line $(\lambda_e+\lambda'_e,\lambda_m+\lambda'_m)$ is  allowed.
For every $G$ there is always a pure electric line $(r_e,0)$ with $r_e \in \Lambda_r$
and a pure magnetic line $(0,r_m)$ with $r_m \in \Lambda_{cr}$.

At this level of the discussion we are still considering the electric and magnetic sublattices of the weight lattice
$\Lambda_w$ and of the co-weight lattice $\Lambda_{mw}$. 
We can project these lattices to the charge (under the center of the gauge group)
lattices thanks to the following observation.
Consider a dyonic line $(\lambda_e,\lambda_m)$, it identifies a sublattice of $\Lambda_w$ and of  
$\Lambda_{mw}$.
We can always add to this line a line in the root lattice $\Lambda_r\times\Lambda_{cr}$, specified by
$(p \, r_e, q \, r_m)$ with $p,q \in \mathbb{Z}$. In this way we reach all the points 
of the lattice with the same charge as $(\lambda_e,\lambda_m) \mod |\mathbf{C}|$.
This explains why the lines can be organized in classes distinguished by their charge under the center.
A generic point is of the form $(e,m) \in \mathbf{C} \times \mathbf{C}$.
A theory is specified by the complete set of allowed charges. 
Not all pairs
of lines are admissible in the quantum theory  due to a
mutual locality condition. 
Take two dyons \((e,m)\) and \((e',m')\); if we fix the position of one and  have the other describe a closed curve around it, the wavefunction picks up a phase proportional to their invariant pairing (\((em' - e'm)\) for \(su(N)\)). Mutual locality implies that this phase must vanish: this is obtained by imposing the \ac{dsz} quantization condition.

\subsection{su(N) lattices and relation to M--theory}

In this section we study the lattices of $su(N)$ and observe that they coincide with the ones obtained
from the M--theory construction. 

When we consider an $su(N)$ algebra, the center is $\mathbf{C}=\mathbb{Z}_N$ 
and, if $N$ is not prime, a generic subgroup  is 
$\mathbf{H} = \mathbb{Z}_{k'}$ where $k' k =N$ and $k,k',N \in \mathbb{N}$.
Since we are interested only in the charges under the center we 
can choose the representatives of the equivalence class of charges.
For example, by observing that the fundamental representation has unitary charge under the center, 
a generic point with charge $(e,m)$ can be associated to the representation
$(Sym^e \framebox[2\width]{~} \,;Sym^m \framebox[2\width]{~})$,
\emph{i.e.} the symmetric product of $e$ and $m$ fundamentals.

In this case, the \ac{dsz} quantization condition for two pairs of charges $(e,m)$ and $(e',m')$ is
\begin{equation}
  \label{DSZSU}
  e m' - m e' = 0 \mod  N,
\end{equation}
as derived in~\eqref{eq:dzs}. 
By following the discussion above, the gauge group in this case can be written
in general as $SU(N)/\mathbb{Z}_k$ (here we also allow  $k=1$ and $k=N$).
The generators of the lattice are fixed to be \((k,0)\) and \((i, k')\) by the fact that 
\(\pi_1(G) = \mathbb{Z}_{k'}\).
The gauge group associated to such a choice is called  $(SU(N)/\mathbb{Z}_k)_i$, and it corresponds 
to the torus $T_{N;k,i}$ defined in the M--theory setup.

We have obtained all the possible lattices in \(\setZ_N \times \setZ_N\): they are generated by the two vectors
$(k,0)$ and $(i,k')$.
This is the crucial observation: these generators coincide with the generators of the lattice \(\Gamma_{N;k,i} \subset \setZ^2\) that we have obtained from M--theory. The other charges are obtained in the field theory construction by combining these vectors
or equivalently by applying the \ac{dsz} condition. In the field theory interpretation this is a condition that 
is intrisically quantum and has to be imposed. In the M--theory interpretation this condition is a classical constraint, and it 
corresponds to the requirement that the \M2--branes are closed curves on the torus.
The quantum condition becomes thus a classical condition on the M--theory side.

We conclude this section with a observation related to the choice of a maximal set of allowed line operators that one has to enforce 
on the field theory side. One may wonder 
whether a theory  containing both \acp{wline} and \acp{hline} lines
only in the adjoint class is consistent.
In field theory nothing appears to forbid this choice. In the M--theory 
derivation this choice is automatically inconsistent because it correspond to an $N^2$ and not to an $N$--covering of the torus.

\subsection{An example: SU(4)}
\label{sec:example}

In this section we study in detail an explicit example. We consider the case of $4$ \M5--branes 
and show how it reduces to $SU(4)$ \ac{sym}.
The possible sublattices associated to the \M5--branes are shown in Figure~\ref{fig:4-M5-lattices}.
The geodesic \M2--branes  probing these sublattices correspond to all the lines connecting the origin with each point of each sublattice (see \emph{e.g.} Figure~\ref{fig:5-2-dyon}).

Once we reduce to \tIIB, these lines become bound states of semi-infinite \F1/\D1 strings. 
We can consider only those connected to the $N$-th brane in the stack. It means that a $(e,m)$ 
bound state is associated to a \ac{wline} in the $e$-index symmetric representation and to an \ac{hline} in the 
$m$-index symmetric representation. This choice is enough to specify the charge lattice.

For example the sublattice \(\Gamma_{4;1,0}\) on top of Figure~\ref{fig:4-M5-lattices} 
defines the torus $T_{4;1,0}$.
Since the generators are \(\langle x_1 = (1,0), x_2 = (0,4) \rangle\), when reducing to \tIIB it corresponds to choosing a pure electric \ac{wline} with charge $(1,0)$ and a pure magnetic \ac{hline} with charge $(0,4)$. In terms of associated representations, the  Wilson lines are in the fundamental class and the \acp{hline} are in the adjoint class. The corresponding gauge group is $SU(N)$.

The $T_{4;2,i}$  tori are associated to a pure electric 
\ac{wline} with charge $(2,0)$. In this case the gauge group is obtained by modding out by a subgroup $\mathbb{Z}_2$ of the center $\mathbf{C}=\mathbb{Z}_4$. The magnetic \ac{hline} $(i,2)$ completely fixes the gauge group. In the first case it has charge $(0,2)$ and the gauge group is $(SU(4)/\mathbb{Z}_2)_0$ while in the second case its charge is $(1,2)$ and the gauge group is $(SU(4)/\mathbb{Z}_2)_1$.
 
The four tori $T_{4;1,i}$ are all associated to \acp{wline} in the adjoint class. Only the purely electric representations invariant under the $\mathbb{Z}_4$ center are admitted. The gauge group is $SU(4)/\mathbb{Z}_4$. The magnetic line is of the form $(i,1)$ where $i=0,1,2,3$ distinguishes the four lattices and fixes the gauge group as $(SU(4)/\mathbb{Z}_4)_i$.

Observe that in this case the number of \M5--branes is not square-free (\(4 = 1 \times 2^2\)) and we have two separate orbits in the S--duality group. At the geometric level, by applying the modular transformation on the $T_{4;k,i}$ 
tori, we observe that $T_{4;2,0}$  is self-dual while the others transform among themselves.
Equivalently, the $(SU(4)/\mathbb{Z}_2)_0$ theory corresponds to a separate orbit of the S-duality group.

\section{Conclusions and open questions}\label{sec:Open}

The connection between the curves probing the torus (or more generally a Riemann surface 
for $\mathcal{N}=2$ theories) and the charge lattices of \textsc{wh} dyons has been discussed 
in the literature~\cite{Drukker:2009tz}.
It has been observed that the global properties are associated to the holonomies of the geodesic curves probing the Riemann surface.
Other proposals for the origin of the global properties from higher 
dimensional theories have been proposed in~\cite{Tachikawa:2013hya,Mayrhofer:2014opa}.

Our analysis is conceptually different. 
We have given a purely geometric prescription, based on M--theory, to fix the global properties
without referring to any specific probing.
The prescription can be summarized as follows:
the gauge group is specified by the choice of  the torus $T_{N;k,i}$ wrapped by $N$ \M5--branes.
This procedure indeed defines the lattice of charges. 
Only as a second step we have probed the lattice by considering \M2--branes
wrapping geodesic curves on $T_{N;k,i}$.
By reducing to \tIIB, we have identified these lattices with the lattices of charges 
of semi-infinite \F1/\D1 bound states probing a stack of $N$ \D3--branes. 
This is the string theory representation of $\mathcal{N}=4$ \ac{sym} with gauge algebra $su(N)$. The 
lattices are further associated to the lattices of charges of \textsc{wh} dyonic states in field theory.
As discussed in~\cite{Aharony:2013hda}, the global properties in this case are associated to the choice of
the $\theta$-angle, and the orbits in the modular group are associated to 
new discrete $\theta$-like parameters.

\bigskip

We would like to stress once more that the classical M--theory properties of our realization of the wrapping imply automatically the conditions that one has to impose in gauge theory:
\begin{itemize}
\item the charges have to satisfy a \ac{dsz} quantization condition;
\item the charge lattice is automatically Lagrangian in \(\setZ_N \times \setZ_N\) (which explains why it cannot be generated by two vectors in the corresponding adjoint classes).
\end{itemize}
Ultimately, the reason why classical properties in M--theory translate into quantum properties in field theory is that we are using only the topology of the torus. After the reduction, topological properties become quantum conditions as they are independent of the gauge coupling $g=R_{10}/R_9$. 

\bigskip

There are many possible extensions of our work.
The global properties associated to the choice of the gauge group are completely geometrized in the 
M--theory description: they are fixed by the torus $T_{N;k,i}$. When reducing to \tIIB these conditions 
should translate to a condition on the RR and NS fluxes carried by the \D3--branes.
Physically one can think of the \D1s which are free to move on the \D3s as generating an electric field 
on them. 
This would imply a quantization condition on the \F1s sourced by the \D3--branes.
It would be interesting to work out this condition in the \tIIB setup and to study how it translates
 into field theory in terms of discrete $\theta$-like terms. 

One can also study systems with reduced supersymmetry. In the $\mathcal{N}=2$ case,
the role of the torus is played by a Riemann surface and also in this case the connection
between the lines and the charge lattices is known~\cite{Drukker:2009tz}.
We expect also here the global properties to be fixed by the embedding of the \M5--branes in the geometry.
One can further reduce to $\mathcal{N}=1$ \ac{sym} and consider the effect of (fundamental) matter fields.

Another aspect that deserves some investigation is the study of the global properties of the symplectic and orthogonal gauge group. At the field theory level, if the rank is large enough, there are in each case two or more
orbits under the action of S--duality. In this case there are non-trivial consequences also in the analysis of Seiberg duality because, as first observed in~\cite{Strassler:1997fe} and then remarked in~\cite{Aharony:2013kma}, in this case 
$Spin(N)$ gauge theories can be dual to $SO(N)$ gauge theories.
In the \tIIB description, theories with symplectic and orthogonal gauge groups are obtained by adding orientifold planes. In the M--theory  description, the projection is more complicated, because of the absence of an open string description of M--theory branes.
Nevertheless we expect that the global properties are fixed by the geometric properties also in this case.
We expect that the geometrization of the action of the \O{}--planes in M--theory done in~\cite{Hanany:1999jy,Hanany:2000fq}
should allow us to recover the charge lattices in a similar manner we did here.
It would be also interesting to study the consequences on the 
brane engineering of~\cite{Amariti:2015yea,Amariti:2015mva} for the
reduction of 4d Seiberg duality to 3d
discussed in~\cite{Aharony:2013dha,Aharony:2013kma}.

We conclude with a comment on AdS/CFT. The different embeddings of the \M5--branes reduce to configurations of multiple \D3s sourcing different combinations of fluxes. Even though we cannot start from the fully back-reacted M--theory solution and simply reduce it to \tIIB (because of the T--duality step), we conjecture that the AdS/CFT duals of the theories with gauge algebra $su(N)$ are the \tIIB configuration with fixed $AdS_5 \times S^5$ geometry sustained by different 3-form fluxes. For example $AdS_5 \times S^5$ with pure NS flux is the dual of \((SU(N)/\setZ_N)_0\).

\section*{Acknowledgments}
\begin{small}
  The authors would like to thank Diego Redigolo for collaboration at an early stage of this project and Noppadol Mekareeya, Jan Troost and  Alberto Zaffaroni for discussions and comments.
  
  \noindent
   A.A. would like to thank \textsc{ccny}, \textsc{ucsd}, Milano-Bicocca and Bern University for hospitality during various stages of this work.

   \noindent
   A.A. is funded by the European Research Council (\textsc{erc}-2012-\textsc{adg}\_20120216) and acknowledges support by \textsc{anr} grant 13-\textsc{bs}05-0001. C.K. acknowledges support by \textsc{anr} grant 12-\textsc{bs}05-003-01 and by Enhanced Eurotalents, which is co-funded by \textsc{cea} and the European Commission. The work of S.R. is supported by the Swiss National Science Foundation (\textsc{snf}) under grant number \textsc{pp}00\textsc{p}2\_157571/1.
\end{small}

\newpage
\appendix

\section{Counting formula}\label{app:Counting}

\subsection{Some facts about (multiplicative) sequences}

\begin{definition}
  A sequence $f$ is multiplicative if
  \begin{equation}
    f( n m ) = f(n) f(m) \, , \quad \text{when $\gcd(n,m) = 1$} .
  \end{equation}
\end{definition}
It follows that $f$ is completely determined by its
values for primes and their
powers, since for any $N$ we can use the factorization $ n = p_1^{k_1}
p_2^{a_2} \dots p_r^{k_r} $ , and
\begin{equation}
  f (N) = f(p_1^{k_1}) f( p_2^{k_2}) \dots f(p_r^{k_r}) \, . 
\end{equation}
Multiplicative sequences form a group under the \emph{Dirichlet convolution}. 
\begin{definition}
  The \emph{Dirichlet convolution} of two sequences $g$ and $h$ is the
  sequence $f$ defined by
  \begin{equation}
    f(n) = ( g * h ) (n) = \sum_{m|n} g(m)\, h ( \frac{n}{m} ) \, ,
  \end{equation}
  where the notation $m|n$ means that the sum runs over all the divisors $m$ of $n$.
\end{definition}
This convolution is commutative, $g*h = h*g$, and
associative, $f * \left( g* h \right) = \left( f * g \right) * h$, and
that the sequence $id $ defined by
\begin{equation}
  id (n) = \set{ 1, 0, 0, \dots } \, 
\end{equation}
is the identity, $f * id = f$.

The information contained in a sequence $f$ can be encoded
into two types of generating functions:
\begin{enumerate}
\item the formal power series (partition function)
  \begin{equation}
   \mathcal{F}(t) = \sum_{n=1}^\infty f(n) t^n \, ; 
  \end{equation}
\item the Dirichlet series
  \begin{equation}
    \mathscr{F}(s) = \sum_{n=1}^\infty \frac{f(n)}{n^s} \, .   
  \end{equation}
\end{enumerate}

If $f$ is multiplicative, its Dirichlet series can be
expanded in terms of an infinite product over the primes, the
\emph{Euler product}:
\begin{equation}
  F(s) = \sum_{n=1}^\infty \frac{f(n)}{n^s} = \prod_{p} \left( 1 +
    \frac{f(p)}{p^s} + \frac{f (p^2)}{p^{2s}} + \dots  \right) \, .
\end{equation}

Both types of generating functions have a simple behavior under
Dirichlet convolution. Let $f, g $ and $h$ be such that
\begin{equation}
 f = g * h \, . 
\end{equation}
The power series for $f$ reads:
\begin{equation}
 \mathcal{F}(t) = \sum_{m=1}^\infty g(m) \mathcal{H}(t^m) = \sum_{k=1}^\infty h(k) \mathcal{G}(t^k)
 \, ,
\end{equation}
and the Dirichlet series is decomposed as
\begin{equation}
  \mathscr{F}(s) = \mathscr{G}(s) \mathscr{H}(s) \, .
\end{equation}

The asymptotic behavior of a sequence can be derived by looking at the
corresponding Dirichlet series. 
\begin{theorem}
  Let $\mathscr{F}(s)$ be a Dirichlet series with non--negative coefficients
  that converges for $\Re(s) > \alpha > 0$, and suppose that $F(s)$ is
  holomorphic in all points of the line $\Re(s) = \alpha $, except for
  $s = \alpha $. If for $s \to \alpha^+$, the Dirichlet series behaves
  as
  \begin{equation}
    \mathscr{F}(s) \sim A(s) + \frac{B(s)}{\left(s - \alpha  \right)^{m+1}} \, ,
  \end{equation}
  where $m \in \setN$, and both $A(s) $ and $B(s) $ are holomorphic in
  $s = \alpha $, then the partial sum of the coefficients is
  asymptotic to:
  \begin{equation}
    \sum_{n = 1 }^N f(n) \sim \frac{B(\alpha )}{\alpha \, m!}
    N^\alpha \log^m (N) \, .
  \end{equation}
\end{theorem}
In order to apply this theorem to our example, we make use of the facts that the Riemann zeta function $\zeta(s)$ is analytic everywhere, except for a simple pole at $s=1$ with residue $1$.

\subsection{A proof for the counting formula}

In this appendix we show that the number of orbits of \(SL(2, \setZ)\) in the set \(\Gamma(N)\) of sublattices of index \(N\) is
\begin{equation}
  \hat f(N) = \sum_{ab^2} 1 
\end{equation}
[\href{https://oeis.org/A046951}{\textsc{oeis a046951}}].\\
A lattice \(\Gamma_{N;k,i} \in \Gamma(N)\) can be equivalently seen as a sublattice of \(\setZ_N \times \setZ_N\). By the Chinese remainder theorem, if \(\gcd(N_1, N_2) = 1\) then \(\mathbb{Z}_{N_1} \times \mathbb{Z}_{N_2} = \mathbb{Z}_{N_1 N_2}\). It follows that \(\hat f(N_1) \hat f(N_2) = \hat f(N_1 N_2)\), \emph{i.e.} that \(\hat f\) is multiplicative. It follows that we only need to prove that
\begin{equation}
  \hat f(p^n) = \left\lfloor \frac{n}{2} \right\rfloor + 1.
\end{equation}
Let \(B\) be the matrix \(B = \begin{psmallmatrix} k & 0 \\ i & k' \end{psmallmatrix} \). There are two actions of \(A \in SL(2,\mathbb{Z})\) on the corresponding lattice \(\Gamma_{N;k,i}\):
\begin{itemize}
\item the action on the left: \( B \mapsto A B\) gives a different basis for the same lattice: \(\Gamma_{AB} = \Gamma_B\);
\item the action on the right: \(B \mapsto B A \) gives the (in general) new lattice generated by the transformed vectors \(\langle A x_1, A x_2 \rangle\).
\end{itemize}
Let \( B \in \Gamma(p^n)\) (the set of lattices of index \(p^n\)).  \(B\) has the form
\begin{equation}
 B = \begin{pmatrix} p^{j_1} & 0 \\ i & p^{j_2} \end{pmatrix},
\end{equation}
where \(0 \le j_i, j_2 \le n \), \(j_1 + j_2 = n\) and \(0 \le i < p^{j_1}\).
If \(i \neq 0\) and \(\gcd(i,p) = 1\), then the matrices of the form \(\begin{psmallmatrix} p^{j_1} & 0 \\ i & p^{j_2} \end{psmallmatrix}\) can be related by an \(S \)--transformation to \(\begin{psmallmatrix} p^{j_1 + j_2} & 0 \\ \lambda p^{j_2} & 1 \end{psmallmatrix}\):
\begin{equation}
  \begin{pmatrix} p^{j_1} & 0 \\ i & p^{j_2} \end{pmatrix} \cdot S = \begin{pmatrix} 0 & p^{j_1}  \\ - p^{j_2} & i\end{pmatrix} = \begin{pmatrix} - \lambda & p^{j_1} \\ - \frac{i \lambda + 1}{p^{j_1}} & i \end{pmatrix} \begin{pmatrix} p^{j_1+j_2} & 0 \\ \lambda p^{j_2} & 1 \end{pmatrix}.
\end{equation}
This is an allowed transformation if \(\lambda \) is the solution of the equation \( i \lambda + 1 = 0 \mod p^{j_1}\). The equation admits one solution if and only if \( \gcd(p^{j_1}, i) = 1\). 
All the matrices of the form \( \begin{psmallmatrix} p^{j_1+j_2} & 0 \\ i & 1 \end{psmallmatrix} \) are in the same orbit because they are related by a T--transformation:
\begin{equation}
  \begin{pmatrix} p^{j_1+j_2} & 0 \\ i & 1 \end{pmatrix} \cdot T = \begin{pmatrix} p^{j_1+j_2} & 0 \\ i + 1 & 1 \end{pmatrix}.
\end{equation}
This gives us the first orbit, identified by the representative
\begin{equation}
  O_0 = \begin{pmatrix} p^{j_1 + j_2} & 0 \\ 0 & 1 \end{pmatrix} = \begin{pmatrix} p^n & 0 \\ 0 & 1 \end{pmatrix}.
\end{equation}
Now we are left with all the matrices of the form \(\begin{psmallmatrix} p^{j_1} & 0 \\ i' p & p^{j_2}\end{psmallmatrix}\), where \(1 \le j_1, j_2 \le n - 1\), \(j_1 + j_2 = n\), \(0 \le i' < p^{j_1 - 1}\). These can be recast into the form \( p \begin{psmallmatrix} p^{j_1 - 1} & 0 \\ i' & p^{j_2 - 1} \end{psmallmatrix}\) so that we can repeat the same reasoning to find a new orbit, with representative
\begin{equation}
 O_1 = p \begin{pmatrix} p^{j_1 + j_2 - 2} & 0 \\ 0 & 1 \end{pmatrix} = \begin{pmatrix} p^{j_1 + j_2 - 1} & 0 \\ 0 & p \end{pmatrix} = \begin{pmatrix} p^{n-1} & 0 \\ 0 & p \end{pmatrix}.
\end{equation}
Now we are left with the matrices of the form \( p \begin{psmallmatrix} p^{j_1 - 1} & 0 \\ i'' p & p^{j_2-1}\end{psmallmatrix} = p^2 \begin{psmallmatrix} p^{j_1 - 2} & 0 \\ i''  & p^{j_2-2}\end{psmallmatrix}\), where \(2 \le j_1, j_2, \le n - 2\), \(j_1 + j_2 = n\), \(0 \le i'' < p^{j_1 - 2}\) and the procedure can be repeated recursively until all \(\Gamma(p^n)\) is exhausted. We obtain a set of orbits with representatives
\begin{equation}
  \left\{ \begin{pmatrix} p^{n-j} & 0 \\ 0 & p^j \end{pmatrix} \right\}_{j=0}^n.
\end{equation}
These orbits are not disjoint since an \(S\)--transformation links them in pairs \(S : O_{j} \mapsto O_{n-j}\). It follows that there are precisely \( \hat f(p^n) = \lfloor \frac{n}{2} \rfloor + 1\) orbits. \(\blacksquare\)

\bigskip

From the expression \(f(p^n) = \lfloor \frac{n}{2}  \rfloor + 1 \) we can immediately write the corresponding Dirichlet series:
\begin{equation}
  \mathscr{F}(s) = \prod_p \left[ 1 + \sum_{n=1}^\infty \frac{f(p^n)}{p^{ns}} \right]  = \prod_p  \sum_{k=0}^\infty \frac{k+1}{p^{2k}} \left( 1 + \frac{1}{p} \right) = \prod_p \frac{1}{1- p^{-s} - p^{-2s} + p^{-3s}} . 
\end{equation}

\section{Lie algebra and representations from the brane perspective}\label{app:Lie}

In this appendix we associate the representation theory for the  
$su(N)$ algebra to a stack of $N$ parallel \D3--branes in \tIIB 
string theory.

The $su(N)$ algebra is associated to the $A_{N-1}$ series.
The Cartan matrix in this case is 
\begin{equation}
\mathcal{A}_{A_{N-1}} = 
\begin{pmatrix}
 2&-1& 0&\dots&\dots&\dots&0\\
-1& 2&-1&0&\dots&\dots&0\\
 0&-1& 2&-1&0&\dots&0\\
 \dots&\dots&\dots&\dots&\dots&\dots&\dots\\
 0&\dots& \dots&0&-1&2&-1\\
 0&\dots&\dots&\dots&0&-1&2\\
\end{pmatrix}
\end{equation}
This case is simply laced and the roots are $\alpha_i = e_i - e_{i+1}$.

The rows of the Cartan matrix correspond to the Dynkin label of the simple roots
in the weight space.
The simple roots can also be represented by the  Dynkin diagram  
\begin{center}
\begin{tikzpicture}
\begin{scope}[node distance=4em, ch/.style={circle,draw,on chain,inner sep=2pt},chj/.style={ch,join},every path/.style={shorten >=4pt,shorten <=4pt},line width=1pt,baseline=-1ex,start chain,shift={(5,0)}]
        \dnode{\(e_1 - e_2\)}
        \dnode{}
        \dydots
        \dnode{\(e_{N-1} - e_{N}\)}
      \end{scope}
\end{tikzpicture}
\end{center}

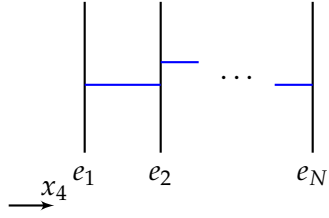
\begin{figure}
  \centering
  \begin{tikzpicture}
    \begin{scope}[thick]
      \foreach \position/\Dname in {-2/{\(e_1\)}, -1/{\(e_2\)}, 1/{\(e_{N}\)}} {
        \draw (\position, 1) -- (\position, -1); \draw (\position,-1.3) node {\Dname}; 
      }
        
      \foreach \leftpos/\rightpos/\heightpos in {-2/-1/-.1, -1/-.5/.2, .5/1/-.1} { 
        \draw[blue] (\leftpos, \heightpos) -- (\rightpos, \heightpos); 
      }
        
      \draw (0,0) node {\(\dots\)};
      
      \tikzhorizaxis{(-2.7,-1.7)}{\(x_4\)}
    \end{scope}
  \end{tikzpicture}  
  \caption{The blue lines connecting the \D3--branes are associated to the \F1--strings (or \D1--branes) and are in 1-1 correspondence with the roots of the algebra, the circles in the Dynkin diagram.}
  \label{fig:Dynkin-branes}
\end{figure}

The brane realization of this system consists of a set of \D1--branes ending on a stack of $N$ \D3--branes. A \D1 between the \(i\)--th and the \(i+1\)--th \D3 corresponds to the root \(\alpha_i\), \emph{i.e.} is a circle in the Dynkin diagram. This system is represented in Figure~\ref{fig:Dynkin-branes}. Any element in the adjoint class (a vector in the adjoint representation) is then in one--to--one correspondence with a set of \D1--branes that have both ends on the stack of \D3s.
It is in fact possible to associate a \D3/\D1 system to any vector in the algebra but in order to do this we will have to use \D1--strings which are infinite in the direction \(x^4\).

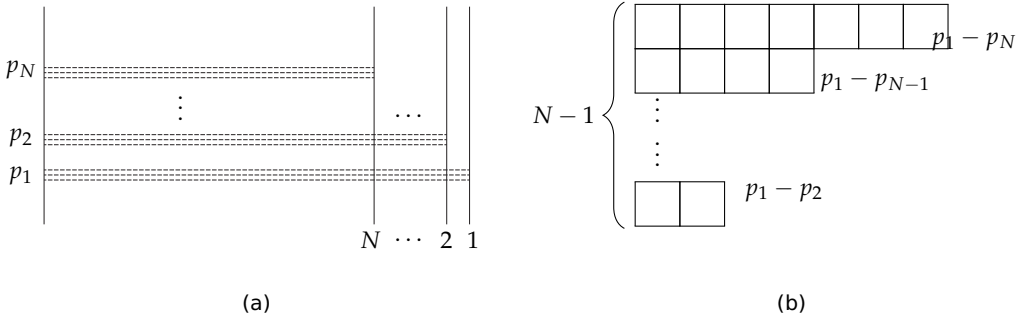
\begin{figure}
  \centering
  \begin{tikzpicture}
    \begin{footnotesize}
    \begin{scope}
      \node at (0,0)  {\includegraphics[width=.4\textwidth]{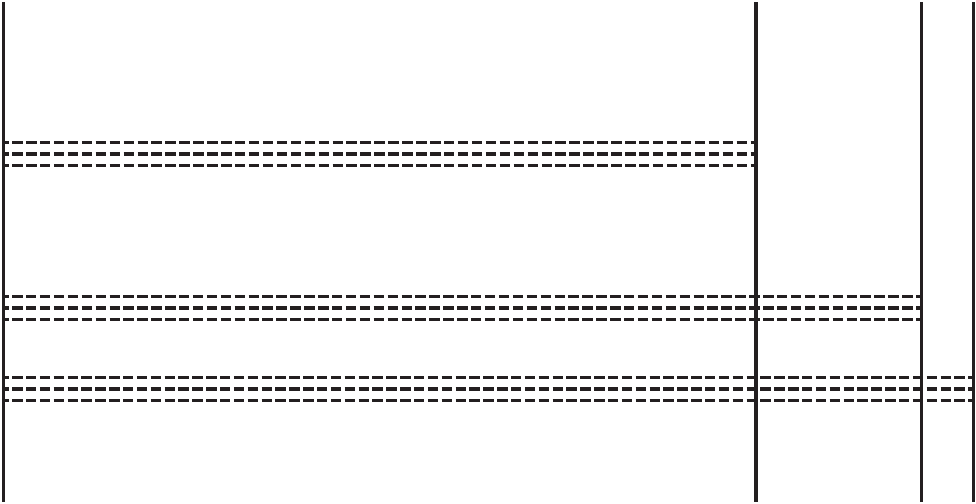}};
      \node at (-3.1,.6) {\(p_N\)};
      \node at (-3.1,-.3) {\(p_2\)};
      \node at (-3.1,-.8) {\(p_1\)};

      \node at (2, 0) {\(\dots\)};
      \node at (-1, .2) {\(\vdots\)};

      \node at (1.5, -1.65) {\(N\)};
      \node at (2, -1.65) {\(\dots\)};
      \node at (2.5, -1.65) {\(2\)};
      \node at (2.85, -1.65) {\(1\)};

      \node at (0, -2.5){\fontfamily{fvs}\selectfont (a)};
    \end{scope}
    \begin{scope}[xshift=.5\textwidth]
      \node at (0,0) {\begin{ytableau}
          {} & {} & {} & {} & {} & {} & {} \\
          {} & {} & {} & {}                \\
          \none[\vdots]                     \\
          \none[\vdots]                     \\
          {}  & {} 
        \end{ytableau}}; 
      \draw [decorate,decoration={brace,amplitude=8pt},xshift=0pt,yshift=0pt]  (-2.2,-1.5) -- (-2.2,1.5) node [black,midway,xshift=-.8cm] {\(N-1\)};
      \node at (2.4, 1) {\(p_1 - p_N\)};
      \node at (1.1, .45) {\(p_1 - p_{N-1}\)};
      \node at (-.1, -1) {\(p_1 - p_2\)};
      \node at (0, -2.5){\fontfamily{fvs}\selectfont (b)};
      
    \end{scope}
    \end{footnotesize}
  \end{tikzpicture}
  \caption{\D3/\D1 configuration (a) and the corresponding Young tableau (b).}
  \label{fig:Young-tableau}
\end{figure}

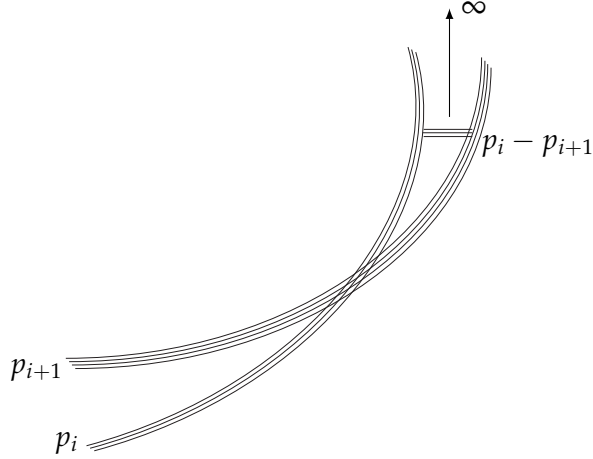
\begin{figure}
  \centering
  \begin{tikzpicture}
    \node at (0,0) {\includegraphics[width=.4\textwidth]{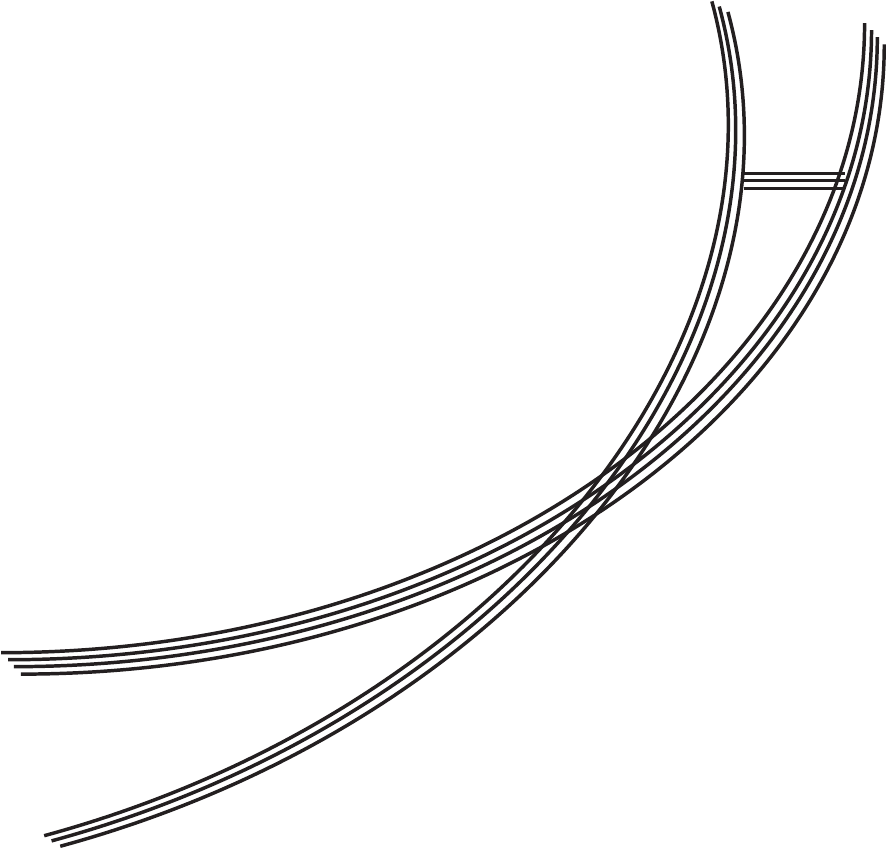}};
    \begin{scope}[x=.8cm, y=.8cm]
      \node at (-4,-2) {\(p_{i+1}\)}; \node at (-3.5,-3.2) {\(p_i\)};

      \node at (4.25,1.75) {\(p_i - p_{i+1}\)};

      \draw [-latex] (2.8,2.2) -- (2.8,4); \node at (3.2,4)
      {\(\infty\)};
    \end{scope}
  \end{tikzpicture}
  \caption{Brane intersection. If \(p_i > p_{i+1}\) the two stacks of \D1--branes intersect. The intersection is solved by emitting \((p_i - p_{i+1})\) that decouple from the system.}
  \label{fig:brane-intersection}
\end{figure}

Start with \(N\) parallel \D3--branes in \(\set{x^0, x^1, x^2, x^3}\). We number them from \(1\) to \(N\) from right to left.
Now we add a set of \D1--branes in \(\set{x^0, x^1}\) and extended in \(x^4\) from one of the \D3s and to \(-\infty\). Let \(p_i\) be the number of \D1s going from \(-\infty\) to the \(i\)-th \D3 (Figure~\ref{fig:Young-tableau}(a)).
This configuration represents a monopole which is in correspondence with a vector \(v \in A_{N-1}\) with weight 
\begin{equation}
  w =  (p_{N-1} - p_N, p_{N-2} - p_{N-1}, \dots, p_1 - p_2) .
\end{equation}
If the weights are all positive (\emph{i.e.} \(p_i \ge p_{i+1}\)), this is the highest weight of a representation with a Young diagram of \(N - 1\) rows (Figure~\ref{fig:Young-tableau}(b)):
\begin{equation} 
  Y = [p_1 - p_N, p_1 - p_{N-1}, \dots, p_1 - p_2 ].
\end{equation}
If \(p_i \ge p_{i + 1}\), using the conventions of~\cite{Moore:2014gua}, we have a configuration in which the \(i\)--th and the \((i+1)\)--st stack intersect in \((p_i - p_{i+1})\) points. If follows that \((p_i - p_{i+1})\) \D1--branes can be created at the intersection and sent to infinity (monopole extraction) (Figure~\ref{fig:brane-intersection}).
For each of these new \D1--branes we obtain a new configuration where the \(i \)--th stack has one brane less and the \((i+1)\)-st one more. The weight of the corresponding vector \(v'\) is \(w' = w - \alpha_i\), \emph{i.e.} \(v'\) is a descendant of \(v\).
In the final configuration where \(p_i - p_{i+1}\) \D1--branes have decoupled, we find that the numerical values of \(p_i\) and \(p_{i+1}\) are interchanged. Repeating the procedure we reach the lowest weight vector of the representation where now the number of branes are \(\set{ p'_1, p'_2, \dots, p'_N} = \set{ p_N, p_{N-1}, \dots, p_1 }\). By construction, \(p'_i \le p'_{i+1}\) and this is a configuration of non-intersecting branes. We find that the monopole extraction reproduces the theory of highest weight representation of the algebra \(A_{N-1}\).

\bigskip

Consider now a generic configuration of \D1--branes ending on a stack of \(N\) \D3s labeled by \(P = \set{ p_1, p_2, \dots, p_N }\). To this we associate a vector \(v \in A_{N-1}\). In general, \(v\) belongs to an infinite number of representations.
There is a natural way (from the point of view of branes) to select a special highest weight representation. 
Recall that if \(p_i > p_{i+1} \) then the \(i\)--th and \(i+1\)--st stack intersect \(p_i - p_{i+1}\) times.
We resolve the intersection by emitting \(p_i - p_{i+1}\) \D1--branes and passing to a new configuration where the numerical values of \(i\) and \(i+1\) are interchanged (this corresponds precisely to subtracting the root \(\alpha_{N-i}\) from \(v\)).
Repeating this procedure one reaches a configuration \(P^{lw}\) where the \(p^{lw}_i\) have the same numerical values but are ordered such that \( p^{lw}_i \le p^{lw}_{i+1}\).
 \(P^{lw}\) corresponds to the lowest weight of the representation identified by the highest weight corresponding to the configuration \(P^{hw}\) where again the \(p^{hw}_i\) have the same numerical values but are ordered such that \(p^{hw}_i \ge p^{hw}_{i+1}\).
This lowest weight state has no intersections at all.
Now we can associate a Young diagram \(Y\) to the initial configuration \(P^0\), namely the Young diagram corresponding to \(P^{hw}\):
\begin{equation}
Y = [p^{hw}_1 - p^{hw}_N, p^{hw}_1 - p^{hw}_{N-1}, \dots, p^{hw}_1 - p^{hw}_2] .
\end{equation}
It is immediate to see that the number of boxes of \(Y\) is given by
\begin{equation}
|Y| = N p^{hw}_1 - \sum_{i=1}^N p^{hw}_i.
\end{equation}
In terms of the configuration \(P\), this number is given by
\begin{equation} 
  |Y(P)| = N \max \set{p_i} - \sum_{i=1}^N p_i = N \max \set{p_i} - m 
\end{equation}
If the \D3--branes in the stack are coincident, we can always add at zero cost any number of \D1--branes between them. This operation will not change \(m\) but can change the value of \(p_1\): by creating/annihilating branes as above we can change \(p_1 \) which can take values between \(0\) and \(N\). It follows that any new configuration \(P'\) obtained in this way will have the number of boxes
\begin{equation} 
|Y(P')|  = N l - m ,
\end{equation}
where \(l \) is an integer \(l \in [0, \dots, N]\). It is always possible to set \(l = 0\) so that the charge of the associated 
\ac{hline} is given by the total number of semi-infinite \D1--branes. Finally we can choose a representative where all the \D1s end on the \(N\)-th \D3, thus picking the highest weight in the \(m\)-index symmetric representation.

\printbibliography

\end{document}